\documentclass{svproc}
\usepackage{url}

\usepackage[hidelinks]{hyperref}
\usepackage{graphicx}
\usepackage{booktabs}
\usepackage{amsmath}
\usepackage{amssymb}

\begin{document}
\mainmatter              

\title{On the Efficiency of Sequentially Aware Recommender Systems: Cotten4Rec}

\author{Shankar Veludandi, Gulrukh Kurdistan, Uzma Mushtaque}

%
%
%
%
%
\institute{Rensselaer Polytechnic Institute (RPI), Troy, NY 12180, USA}

\maketitle              

\begin{abstract}
Sequential recommendation (SR) models predict a user's next interaction by modeling their historical behaviors. Transformer-based SR methods, notably BERT4Rec, effectively capture these patterns but incur significant computational overhead due to extensive intermediate computations associated with Softmax-based attention. We propose Cotten4Rec, a novel SR model utilizing linear-time cosine similarity attention, implemented through a single optimized compute unified device architecture (CUDA) kernel. By minimizing intermediate buffers and kernel-launch overhead, Cotten4Rec substantially reduces resource usage compared to BERT4Rec and the linear-attention baseline, LinRec, especially for datasets with moderate sequence lengths and vocabulary sizes. Evaluations across three benchmark datasets confirm that Cotten4Rec achieves considerable reductions in memory and runtime with minimal compromise in recommendation accuracy, demonstrating Cotten4Rec's viability as an efficient alternative for practical, large-scale sequential recommendation scenarios where computational resources are critical.\footnote{Code and documentation are available at \href{https://github.com/shankar-veludandi/Portfolio/tree/main/cotten4rec}{https://github.com/cotten4rec}-Accepted to Intellysis 2026
}
\end{abstract}
\keywords{recommender systems, efficient transformers, attention mechanism, sequential recommender systems, cosine attention}
\section{Introduction}
The next-item prediction (NIP) task involves predicting the next relevant item(s) that users will interact with. This task falls into the category of the SR problem and applies to e-commerce, social media, and subscription platforms ~\cite{zhao_2021} ~\cite{liu_2023} ~\cite{yang2019pretrainingcontextawareitemrepresentation}. The dynamic semantic context between items selected by a user can help solve the NIP task because these interactions contain valuable information~\cite{8594844}. Transformer models~\cite{vaswani2023attentionneed} help capture this context through a self-attention mechanism and have been a building block for various sequential models for the NIP task, such as BERT4Rec~\cite{sun2019bert4recsequentialrecommendationbidirectional}and SASRec~\cite{kang2018selfattentivesequentialrecommendation}. Most Transformer-based models are built on the idea of well-known LLMs such as BERT~\cite{devlin2019bertpretrainingdeepbidirectional} or GPT~\cite{openai2024gpt4}. In recommender systems (RecSys) models, an item selected by a user replaces the word in the LLM sequence. 

Sequential recommender systems (SRSs) have gained traction due to their ability to model evolving user preferences. A variety of approaches have emerged to improve recommendation quality, such as contrastive learning, memory-based methods, and hierarchical models. However, most of these methods focus heavily on accuracy while underemphasizing computational efficiency. Many of them rely on language modeling ideas, yet key differences exist. Unlike natural language, user interaction sequences may be much longer and drawn from dynamic item vocabularies that grow over time, presenting scalability challenges.

A significant portion of SR research relies on the idea of language modeling. There are similarities between the SR problem and language modeling, but there are also significant differences. A sentence is a finite-length text. The NIP task, on the other hand, can involve an entire user's selection history, which can lead to very long sequences. Language models comprise words from a specific language, which are finite in number. SR problem witnesses an inclusion of new items daily, thus adding to the diversity of tokens (item representations), making it a computationally challenging task. Due to these reasons, there is a need for improved efficiency (in terms of time and memory) and better scalability for RecSys models, particularly for the SR problem.

Transformer-based models employ self-attention mechanisms~\cite{vaswani2023attentionneed} that learn sequence representations via a dot-product operation between three matrices. This operation captures item relationships and context fairly well, as shown by experimental studies~\cite{sun2019bert4recsequentialrecommendationbidirectional}. The computational and memory complexity of this model is quadratic in both the sequence length (N) and the model (embedding) dimension (d) . This can be challenging for NIP due to longer sequences and new item additions involved. Many techniques address the computational and memory efficiency issue in language models~\cite{zhu2021longshorttransformerefficienttransformers},~\cite{wang2020linformerselfattentionlinearcomplexity}, ~\cite{beltagy2020longformer} and computer vision models~\cite{10.1007/978-3-031-25082-8_3}. These models have been adapted to build a vast body of research that utilizes linear attention models in the RecSys domain to solve the NIP task efficiently ~\cite{liu_2023}. Most of these models focus on reducing time complexity through low-rank approximation of input matrices, but they still have a large number of model parameters and a high memory footprint. In this work, we introduce a computationally efficient RecSys model called Cotten4Rec, which utilizes cosine attention (Cotten) instead of dot-product attention. Cosine attention has been applied in other domains such as text~\cite{mongaras2024cottentionlineartransformerscosine} and visual representations in vision transformers~\cite{10549904} ~\cite{LI2024108386}. To the best of our knowledge, this is the first work to utilize cosine attention in the RecSys domain for the SR problem. Our proposed model aims to retain the advantages of attention while also reducing memory footprint, computational complexity and increasing scalability. Bidirectional frameworks such as BERT4Rec~\cite{sun2019bert4recsequentialrecommendationbidirectional} have been found to be more effective in preserving the temporal context of sequences. Therefore, we use the BERT4Rec architecture for our experiments. The model itself is generic and can be used with any Transformer based framework.

The key contributions of this work are:

1. We develop a novel RecSys model, Cotten4Rec, which utilizes cosine attention. It is a linear attention model that reduces the computational complexity of $\mathcal{O}(s^2)$ to $\mathcal{O}(s)$, where $s$ is the sequence length. This is one of the first works to utilize cosine attention in the RecSys domain for the SR problem.

2. Cotten4Rec uses an efficient computation of attention weights by decoupling the normalization step from the matrix multiplication operation. This enables faster computation by using optimized linear algebra routines.

3. Cotten4Rec differs from other linear kernel-based approximation models used for the SR problem due to its higher numerical stability and direct attention calculation rather than a low-rank approximation.

4. The proposed model can be used with any Transformer-based architecture. In this study, we implement it on a BERT4Rec architecture because of the known advantages of this architecture for the SR problem.

5. Cotten4Rec achieves a $\approx23\%$ lower memory footprint compared to BERT4Rec (SOTA) and LinRec (linear baseline).

6. Cotten4Rec achieves faster training times versus BERT4Rec and LinRec by $\approx4\%$ and $\approx20\%$, respectively.

\section{Related Work}

\subsection{Scaled Dot-Product Attention}
"Attention Is All You Need"~\cite{vaswani2023attentionneed} is a 2017 landmark research paper in machine learning. The paper introduced a new deep learning architecture known as the transformer, based on the attention mechanism proposed in 2014~\cite{bahdanau2016neuralmachinetranslationjointly}. Scaled dot-product attention is a fundamental mechanism in the Transformer architecture~\cite{vaswani2023attentionneed}. It operates on three input matrices: query \( Q \), key \( K \), and value \( V \), each representing different projections of the input data. The attention mechanism computes a similarity score between the query and key matrices using a dot-product, producing an attention weight matrix that determines the relative importance of each key-value pair. To ensure numerical stability and prevent excessively large values, the dot-product is scaled by the square root of the key dimension, \( \sqrt{d} \). The final attention output is obtained by applying a softmax function to normalize the scores and then weighting the value matrix \( V \). Mathematically, it is expressed as:

\begin{equation}
    A(Q, K, V) = \text{softmax} \left(\frac{QK^T}{\sqrt{d}} \right) V.
    \label{eq:vaswanidotproduct}
\end{equation}

where $Q$, $K$, and $V$ represent the query, key, and value matrices, respectively, and $d$ is the dimensionality of the key vectors. This mechanism computes attention weights by taking the dot-product of $Q$ and $K^\top$, scaling the result by $\sqrt{d}$, and applying the softmax function to obtain a probability distribution over the keys. These probabilities are then used to compute a weighted sum of the values $V$. The scaled dot-product attention mechanism enables the model to focus on the most relevant parts of the input sequence for each word, effectively capturing long-range dependencies and enhancing the model's capability to process complex sequential data. The scaling factor $\sqrt{d}$ is crucial for stabilizing training, as it prevents the dot-products from growing too large, which could otherwise cause vanishing gradients or numerical instability in the softmax computation.

\subsection{Recommender System Models for the SR Problem}

BERT4Rec (Bidirectional Encoder Representations from Transformers for Sequential Recommendation) \cite{sun2019bert4recsequentialrecommendationbidirectional} is a transformer-based model designed for sequential recommendation tasks. It adapts the BERT \cite{devlin2019bertpretrainingdeepbidirectional} architecture from NLP to learn sequences of user behaviors in recommender systems.

The BERT4Rec model architecture consists of the following key components:

\textbf{Input Embedding:} Given a user interaction sequence $S = [s_1, s_2, \dots, s_T]$, where $s_t$ represents an item interacted with at timestep $t$, each item is mapped to a dense vector embedding. Additionally, positional encodings $P_t$ are added to preserve order information:
\begin{equation}
E_t = s_t + P_t.
\end{equation}

\textbf{Transformer Encoder:} BERT4Rec uses a stack of self-attention layers to model dependencies between items in a bidirectional manner. Each layer applies the multi-head self-attention mechanism:
\begin{equation}
\text{Attention}(Q, K, V) = \text{softmax}\left( \frac{QK^T}{\sqrt{d_k}} \right)V,
\end{equation}
where $Q$, $K$, and $V$ are query, key, and value matrices derived from input embeddings, and $d_k$ is the dimension of the key vectors.

\textbf{Masked Training Objective}: Unlike autoregressive models, BERT4Rec employs masked language modeling (MLM). Some items in the sequence are randomly masked, and the model is trained to predict them using the surrounding context:
\begin{equation}
\mathcal{L} = -\sum_{i \in \mathcal{M}} \log P(s_i | S_{\backslash i}),
\end{equation}
where $\mathcal{M}$ denotes the set of masked positions, and $S_{\backslash i}$ represents the sequence with masked items removed.

\textbf{Prediction Layer}: The final hidden state of each item is passed through a softmax layer to compute probabilities over the item vocabulary:
\begin{equation}
P(s_i | S) = \text{softmax}(W h_i + b),
\end{equation}
where $h_i$ is the hidden representation of the masked item, and $W$, $b$ are learnable parameters.

BERT4Rec \cite{sun2019bert4recsequentialrecommendationbidirectional} improves upon previous methods in sequential recommendation by leveraging a bidirectional transformer-based architecture. Earlier models, such as GRU4Rec \cite{hidasi2016sessionbasedrecommendationsrecurrentneural}, utilized gated recurrent units (GRUs) but struggled with capturing long-range dependencies effectively. SASRec \cite{kang2018selfattentivesequentialrecommendation}, a unidirectional transformer model, considered only past interactions, limiting its ability to model sequential dependencies fully. In contrast, BERT4Rec captures both past and future dependencies within item sequences through its bidirectional self-attention mechanism. This design enables it to model richer contextual information, leading to improved recommendation accuracy. Moreover, BERT4Rec efficiently handles long sequences by leveraging self-attention, making it highly scalable. 

Here's your draft with additional context on ELU clearly explained, along with its formula:

\subsection{Linear Attention Models}

Most Transformer‐based recommenders, such as BERT4Rec, rely on softmax scaled dot‐product attention, which requires forming an $s \times s$ score matrix and incurs $O(s^2)$ memory and time costs. Linear attention models replace the softmax step with a simpler feature map that computes attention using two smaller matrix multiplications, reducing both memory and computation to $O(s\,d^2)$.

A representative example is LinRec, which applies an element‐wise nonlinear transformation—specifically, the Exponential Linear Unit (ELU)—to the key and query projections. The ELU activation function, defined as:

\[
\text{ELU}(x) = 
\begin{cases}
x & \text{if } x > 0, \\
\alpha(e^{x} - 1) & \text{otherwise},
\end{cases}
\]

introduces nonlinearity while maintaining smooth gradients and avoiding vanishing gradient issues. LinRec specifically uses ELU + 1 to ensure all attention features are positive. By doing so, LinRec avoids explicitly computing the large $s \times s$ attention score matrix: instead, it first collapses transformed keys and values into a compact $d \times d$ "key–value" product and then multiplies that result by the transformed queries. This two‐pass procedure maintains the ability to model pairwise interactions within a user's history while running in linear time and memory relative to sequence length.

Because LinRec never materializes an $s \times s$ buffer, its peak memory usage grows as $O(s\,d + d^2)$ instead of $O(s^2 + s\,d)$. In practice, this allows processing moderately long user sequences without exhausting GPU resources. Moreover, by replacing softmax with a simpler ELU‐based feature map, LinRec requires fewer kernel launches and global memory accesses than traditional softmax‐based implementations.

Subsequent work in both NLP and vision has explored alternative linear maps, such as kernel approximations or randomized projections, but LinRec remains a widely cited baseline in sequential recommendation due to its straightforward architecture and strong empirical performance. ~\cite{katharopoulos2020transformersrnnsfastautoregressive} ~\cite{choromanski2022rethinkingattentionperformers}.

\subsection{Cosine Attention-Based Models}
Cosine similarity is used as a similarity measure in various domains and is defined as a measure of similarity between two vectors in an inner-product space. By definition, cosine similarity between two vectors is the dot-product of two vectors divided by their respective Euclidean norms. Cosine similarity focuses on direction rather than magnitude. Cosine attention utilizes the property of matrix norms to be decoupled by normalizing each matrix row-wise and then performing matrix multiplication \cite{mongaras2024cottentionlineartransformerscosine}. The cosine attention formulation is more efficient compared to the standard attention mechanism and allows for optimized implementation in CUDA.

Time and memory efficiency are important concerns for the SR problem given longer item sequences \cite{LLiu_2023}. In this work, we utilize cosine attention to solve the SR problem more efficiently in the BERT4Rec architecture. We propose a novel model called Cotten4Rec, which generalizes cosine attention applied to the SR system domain. Furthermore, by utilizing a native CUDA kernel, the cosine-similarity-based model (Cotten4Rec) has the potential to be more memory efficient in GPU-accelerated environments.

\section{Methodology}

In this section, we describe the Next item prediction (NIP) problem mathematically, followed by an explanation of the attention mechanisms and the Cotten4Rec model.

\subsection{Problem Statement}
The general recommendation system problem is to predict relevant items from a given set $V=\{v_1,v_2, \dots, v_{|V|}\}$ to a given set of users $U=\{u_1,u_2, \dots, u_{|U|}\}$. In a sequential recommendation (SR) problem, we consider the chronological user-item interaction history to predict the next item a user will select. The interaction sequence varies from user to user. In general, the SR problem involves predicting the item that a user will select at a particular time step. To solve this problem, the conditional probability of selecting this next item (conditioned on the entire previous sequence) is calculated.

\subsection{Attention Mechanisms}
We build on the BERT4Rec framework to investigate three distinct attention approaches for sequential item modeling. First, BERT4Rec employs standard scaled dot‐product attention, which incurs a quadratic $\mathcal{O}(N^2)$ complexity in sequence length. Second, LinRec leverages a linear attention mechanism that replaces softmax with an ELU+1 kernel, reducing memory usage for longer sequences. Finally, Cotten4Rec adopts a cosine-similarity-based attention, featuring row‐wise normalization of $Q$ and $K$ and a learnable scaling factor on $V$. We aim to maintain strong recommendation accuracy while lowering memory and runtime overhead.

\subsection{Model Description}
Cotten4Rec utilizes the BERT4Rec architecture \cite{sun2019bert4recsequentialrecommendationbidirectional}, which is an encoder model built on the Transformer layer with self-attention \cite{vaswani2023attentionneed}. Similar to BERT4Rec, Cotten4Rec has the following components:

\textbf{Input Embedding:} Each user’s interaction history is captured as a sequence of items. Items are represented as learnable embedding vectors with added position embeddings to include sequential order.

\textbf{Transformer Layers and Cosine Attention:} Cotten4Rec uses the Multi-Head Attention (MHA) framework to capture multiple representations of the sequence and combine them. For a given sequence of length $n$ and embedding dimension $d$, we calculate the hidden representation $h_i^l$ for each transformer layer $l$ at each sequence position $i$. These hidden representations are stacked into a matrix $H^l \in \mathbb{R}^{n \times d}$. MHA facilitates the linear projection of $H^l$ into $h$ subspaces having learnable projections. Attention is then applied in parallel to output representations that are concatenated and once again projected. For example:

\begin{equation}
MHA(H^l) = \text{concat}(\text{head}_1, \text{head}_2, \dots, \text{head}_h) W^o
\end{equation}
Here, $W^o$ is the learnable weight matrix.
\begin{equation}
\text{head}_i = \text{CosineAttention}(H^l W_i^Q, H^l W_i^K, H^l W_i^V)
\end{equation}
where $W_i^Q, W_i^K, W_i^V$ are learnable projection matrices. The parameters from these matrices are not shared across layers. The CosineAttention function is defined as:

\begin{equation}
\text{CosineAttention}(Q,K,V) = \text{Sim}(Q,K) \cdot V 
= \frac{Q}{\|Q\|_2} \cdot \frac{K^T}{\|K\|_2} \cdot V
\label{eq:cosineattention}
\end{equation}

In Equation~\eqref{eq:cosineattention}, $Q$, $K$, and $V$ are query, key, and value matrices that represent different projections from the same matrix $H^l$. The cosine similarity is calculated after row-wise $L_2$ normalization of $Q$ and $K$, replacing the SoftMax in standard attention. This enables efficient attention computation with reduced memory cost using linear algebra kernels optimized in CUDA \cite{mongaras2024cottentionlineartransformerscosine}.

To address stability issues, as cosine similarity can yield row-wise sums exceeding 1, we apply a learnable scaling factor $m$:
\begin{equation}
\text{CosineAttention}(Q,K,V) = \frac{1}{n^m} \cdot \text{Sim}(Q,K) \cdot V
\end{equation}

We reorder operations to compute:
\begin{equation}
\text{CosineAttention}(Q,K,V) = \frac{Q}{\|Q\|_2} \cdot \left(\frac{K^T}{\|K\|_2} \cdot V \right)
\end{equation}
This outer product requires memory of order $d^2$ and is efficient for typical SR cases where $n \gg d$. Cotten4Rec is thus a memory-efficient model. Moreover, as shown in \cite{mongaras2024cottentionlineartransformerscosine}, \cite{katharopoulos2020transformersrnnsfastautoregressive}, cosine attention can be viewed as an RNN, resulting in constant memory usage through outer product updates.

```latex
\subsection{Cosine Similarity and CUDA Kernel for Cotten4Rec}
The heart of Cotten4Rec’s efficiency lies in a single, highly‐optimized CUDA kernel (adapted from Mongaras et al. \cite{mongaras2024cottentionlineartransformerscosine}) that fuses
\begin{enumerate}
  \item row‐wise $L_2$ normalization of $K$ and $V$,
  \item accumulation of $(K^T\,V)\in\mathbb{R}^{d\times d}$ entirely in registers/shared memory, and
  \item multiplication of $Q$ by the resulting $d\times d$ matrix,
\end{enumerate}
all before writing the final $n\times d$ “context” matrix back to global memory. The key steps are:

\begin{itemize}
  \item \textbf{Shared‐memory tiling:} We load small $T\times d$ tiles of $K$ and $V$ (for some tile size $T\ll n$) into shared memory or registers. Each warp (or thread‐block) is responsible for one tile pair $(K_{[i:i+T)},\,V_{[i:i+T)})$.
  \item \textbf{On‐the‐fly $L_2$ normalization:} Within each tile, each thread computes the squared $L_2$ norm of its $d$‐dimensional row of $K$ (and similarly for $V$) by reducing $\sum_{j=1}^d K_{ij}^2$. Then it divides the row by $\sqrt{\sum_j K_{ij}^2 + \varepsilon}$, all in registers. This avoids allocating a separate $n\times d$ “normalized” buffer.
  \item \textbf{Accumulate $K^T V$ in registers:} After normalization, each thread pair contributes to multiple entries of the $d\times d$ matrix $(K^T\,V)$ by computing $\sum_{i=i_0}^{i_0+T-1} K_{i\ast}^T\,V_{i\ast}$ over the tile. These partial products are accumulated in thread‐private registers (or, when necessary, in shared memory) so that no intermediate $n\times d$ matrix is ever written to global memory.
  \item \textbf{Final $Q \times (K^T V)$ multiplication:} Once $(K^T\,V)$ is fully reduced across all tiles (i.e.\ summed over $i=1\ldots n$), we launch a fused pass (within the same kernel) that loads $Q$ tile by tile, normalizes each row of $Q$ on the fly, and multiplies by the precomputed $d\times d$ matrix in registers/shared memory. The result—an $n\times d$ “context” matrix—is written directly to global memory.
\end{itemize}

Because all three phases (normalization of $K/V$, accumulation of $(K^{\top}V)$, and application to $Q$) occur within one CUDA kernel launch, we incur only a single kernel-launch latency (approximately 10--20~$\mu$s) and allocate no large intermediate buffers beyond the final $n \times d$ output. This stands in stark contrast to (a) standard Softmax attention, which needs an intermediate $n \times n$ “scores” buffer and multiple kernels ($QK^{\top}$ GEMM, Softmax, final GEMM), and (b) LinRec’s ELU+GEMM pipeline, which allocates $n \times d$ feature maps and launches at least three kernels (ELU, GEMM-1, GEMM-2).

\paragraph{Bidirectional Attention Adaptation.}
We modify the original Mongaras et al.\ kernel—designed for causal (autoregressive) language modeling—so that:
\begin{itemize}
  \item \textbf{No causal masking:} We simply omit any triangular mask since recommendation is bidirectional; each position can attend to all other positions regardless of order.
  \item \textbf{Data‐type restriction:} We support only \texttt{float32} and \texttt{float16} (via AMP) to ensure numerical stability and maximize throughput. All $L_2$‐norms and partial sums are computed in \texttt{float32}, while inputs and outputs may use \texttt{float16} when AMP is enabled.
  \item \textbf{Cumulative sums via warp‐reductions:} To perform the per‐tile partial sums $\sum_i K_{ij}^2$ and $\sum_i K_{ij}\,V_{ik}$ efficiently, we use warp‐intrinsics (\texttt{\_\_shfl\_down\_sync}) or shared‐memory reductions, ensuring that the cost of reduction is $\mathcal{O}(d)$ per tile rather than $\mathcal{O}(n\,d)$ overall.
\end{itemize}

\paragraph{Memory and Compute Complexity.}
\begin{itemize}
  \item \textbf{Memory:} At any point, we only store (i) the $n\times d$ input $Q,K,V$, (ii) the final $n\times d$ output, and (iii) the $d\times d$ accumulator $(K^T\,V)$. Thus peak memory usage for attention is $\mathcal{O}(n\,d + d^2)$, as opposed to $\mathcal{O}(n^2 + n\,d)$ in standard Softmax.
  \item \textbf{Compute:} We perform three $\mathcal{O}(n\,d^2)$ passes: one to build $(K^T\,V)$, one to normalize $Q$ on the fly, and one to multiply $Q$ by $(K^T\,V)$. In total, the arithmetic cost is $\approx 2\,n\,d^2$ (excluding the initial $Q,K,V$ projections), matching LinRec in $\mathcal{O}(n\,d^2)$ complexity. However, because all is fused into one kernel, constant‐factor overhead is cut dramatically.
\end{itemize}

\paragraph{Automatic Mixed Precision (AMP).}
To further accelerate matrix operations, we employ PyTorch’s \texttt{torch.cuda.amp} with \texttt{bfloat16}: 
\begin{itemize}
  \item All reductions (e.g.\ row‐norms and partial dot‐products) are done internally in \texttt{float32}.
  \item Multiplications and final output writes may be in \texttt{bfloat16} to double throughput and halve memory footprint.
\end{itemize}

```

\subsection{Objective Function}
The training objective of Cotten4Rec follows the negative log-likelihood loss used in BERT4Rec \cite{sun2019bert4recsequentialrecommendationbidirectional}. Given a sequence of interactions $S = \{s_1, s_2, \dots, s_T\}$, a subset of positions is randomly masked, and the model is trained to predict the masked items. The probability of predicting the correct item $s_t$ at position $t$ is given by:

\begin{equation}
P(s_t \mid S_{\backslash t}) = \frac{\exp(\text{score}(s_t))}{\sum_{v \in V} \exp(\text{score}(v))}
\end{equation}

where $\text{score}(s_t)$ is the dot-product between the final hidden representation of the masked position and the corresponding item embedding. The loss function is then formulated as:

\begin{equation}
\mathcal{L} = - \sum_{t \in M} \log P(s_t \mid S_{\backslash t})
\end{equation}

where $M$ is the set of masked positions. This objective encourages the model to learn informative and context-aware item embeddings for accurate recommendation.

\section{Model Training}
Building upon the methodological foundations described earlier, we present the overall architecture and training procedure of the Cotten4Rec model. Cotten4Rec integrates a novel \textit{cosine-similarity-based attention mechanism} within a BERT-style framework for sequential item prediction. Each Transformer block consists of multi-head attention (utilizing cosine similarity instead of dot products) and a two-layer feed-forward network (FFN) with GELU activation, connected via residual connections and layer normalization.

We adopt a \textit{masked item prediction} strategy to learn robust user-item representations: a random subset of items in each sequence is masked, and the model is trained to recover these items using only the surrounding context. This enables a bidirectional understanding of user histories, capturing both preceding and subsequent dependencies for each position. The Transformer outputs are then passed through an additional two-layer FFN to yield a probability distribution over the item vocabulary. We minimize the negative log-likelihood of the masked items, promoting accurate sequential recommendations under constrained memory and runtime budgets.

\subsection{Experiments}
We conduct experiments on real-world datasets to answer the following research questions:

\begin{itemize}
    \item \textbf{RQ1:} What is the trade-off between efficiency gains (memory and runtime) and performance of Cotten4Rec compared to SOTA BERT4Rec?
    \item \textbf{RQ2:} How does LinRec, a known linear model, compare to Cotten4Rec?
    \item \textbf{RQ3:} How do changes in sequence length and embedding size affect Cotten4Rec's performance and efficiency?
\end{itemize}

\subsubsection{Task Definition}
We follow a standard \textit{leave-one-out} (next-item) evaluation scheme, a common practice in sequential recommendation. We hold out each user's final (most recent) interaction as the test item while the preceding interactions form the training set. Models are trained to predict this final item given the user's historical sequence, reflecting the real‐world need to recommend the next item a user will select.

\subsubsection{Baselines and Approach}
BERT4Rec is used as the primary baseline, implemented as a standard Transformer using scaled dot-product attention. LinRec, a linear attention-based model, is also included due to its efficiency with long sequences. Cotten4Rec retains the core BERT4Rec components—embedding layers and transformer blocks—but replaces the attention mechanism with cosine similarity via a custom CUDA kernel, aiming to reduce memory and runtime without compromising accuracy.

\subsubsection{Evaluation Metrics}
We evaluate our models using two widely adopted top-$k$ ranking metrics. The first is Normalized Discounted Cumulative Gain (NDCG@10), which rewards placing the correct item higher in the ranked list of 10 items. The second is Hit Ratio (HIT@10), which checks whether the correct item appears in the top 10, regardless of its position. In addition to these accuracy metrics, we also record peak GPU memory usage, epoch-level training time, and the total number of model parameters. These measurements help assess the computational efficiency of each model alongside its recommendation quality.

\subsubsection{Sequence Length and Embedding Size}
We evaluate all models under various configurations of sequence length and embedding size. For sequence lengths, we use 50, 100, and 200 for the MovieLens datasets and 20, 50, and 100 for the Beauty dataset. For embedding sizes, we experiment with values of 64, 128, and 256. These variations are designed to test the models’ scalability with respect to memory consumption, runtime efficiency, and recommendation quality.

\section{Datasets and Preprocessing}
We evaluate the proposed models on three real-world representative datasets chosen for their diversity in domain and scale:

\begin{enumerate}
    \item \textbf{ML-1M (MovieLens 1M):} This widely used dataset contains approximately 1 million user-item interactions in the movie domain. Due to its moderate scale and well-defined rating behaviors, it serves as a popular benchmark for recommender systems.

    \item \textbf{ML-20M (MovieLens 20M):} This larger variant of the MovieLens dataset features 20 million user-item interactions. It offers longer user histories, enabling evaluations of scalability and performance on more extensive sequences. For preprocessing, we filtered users to those with 10 to 200 interactions and randomly sampled 5,000 users for validation.

    \item \textbf{Beauty (Amazon Beauty):} A product-rating dataset focused on beauty items sold on Amazon. Compared to MovieLens, user sequences here are shorter, yet the dataset spans a broad range of product categories, supporting useful generalizability. Preprocessing retained users with 5 to 200 interactions; no further sampling was applied.
\end{enumerate}

Table~\ref{tab:dataset-summary} summarizes key statistics for each dataset, including user and item counts, average sequence length, and dataset density. Overall, this preprocessing strategy yields sufficiently rich sequential data for our experiments while managing the computational overhead associated with long user histories in large-scale domains.

\begin{table*}
  \caption{Summary of datasets including user and item statistics, sequence length, and density.}
  \label{tab:dataset-summary}
  \begin{tabular}{l c c c c c c}
    \toprule
    Dataset & Total \# of Users & \# of Users After Preprocessing & Total \# of Items & Total Interactions & Avg. Sequence Length & Density \\
    \midrule
    ML-1M & 6,040 & 6,040 & 3,706 & 1,000,209 & 166 & 95.53\% \\
    Beauty & 1,210,271 & 52,361 & 120,472 & 466,364 & 9 & 99.99\% \\
    ML-20M & 138,493 & 111,894 & 16,569 & 7,573,787 & 68 & 99.59\% \\
    \bottomrule
  \end{tabular}
\end{table*}

\section{Implementation Details}

\subsection{Hyperparameter Tuning}
We performed grid and sequential searches over a core set of hyperparameters intended to remain fixed across all experiments. Based on preliminary evaluations, we set the learning rate to 0.001, weight decay to 0.001, dropout rate to 0.1, and gradient clipping threshold to 1.0. A batch size of 128 was used throughout.

Model training was conducted for 20 epochs on the MovieLens datasets. For the Amazon Beauty dataset, training was limited to 5 epochs, as the models exhibited faster convergence in NDCG@10 and HIT@10. As a result, we compared training times across datasets by dividing the total training time by the number of epochs. This reduction also helped to manage total runtime without compromising performance. 

All experiments were repeated with three random initializations (seeds 0, 42, and 123) to ensure that our results are robust to stochastic variation.

\subsection{Hardware and Software Environment}

All experiments were executed on a single node equipped with an NVIDIA Tesla V100-SXM2 GPU (32 GB), an Intel Xeon Gold 6248 CPU (2 sockets, 40 cores each), and 755 GB RAM, supported by a 5.8 TB NVMe SSD. We used CUDA 12.2 (Driver 535.104.05) for GPU-accelerated operations. Our software stack included Python 3.9.21 and PyTorch 2.0.0+cu118. Additional libraries, such as TorchVision 0.15.1+cu118, Transformers 4.49.0, and FastAttention2 (for our custom CUDA kernel), were employed to handle model components, data processing, and visualization (via NumPy 1.26.4, Pandas 2.2.3, and Matplotlib 3.9.4). We also utilized the Huggingface Hub (v0.29.2) for optional model management and version control. This setup ensured a consistent experimental environment for all models, enabling us to compare memory usage, training speed, and recommendation accuracy in a meaningful manner.

\section{Results}
\begin{figure*}
  \centering
  \includegraphics[width=\textwidth]{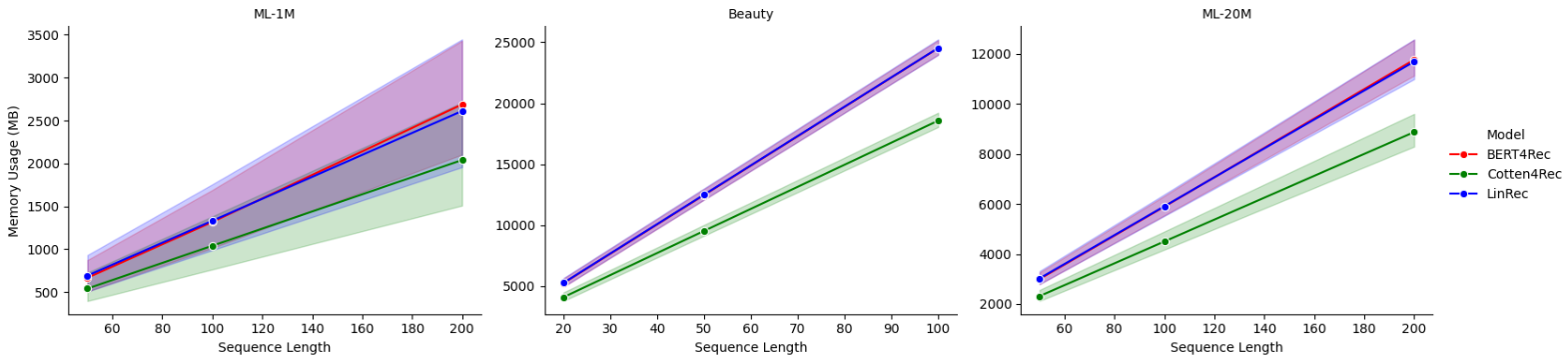}
  \caption{Memory Usage vs Sequence Length}
  \label{fig:MemvsSeqL}
\end{figure*}

\begin{figure*}
  \centering
  \includegraphics[width=\textwidth]{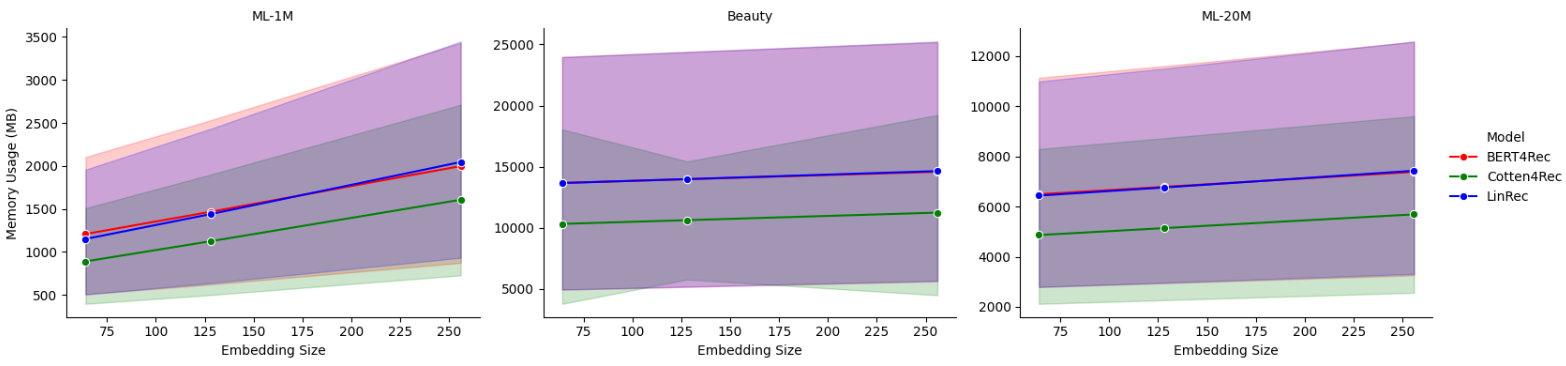}
  \caption{Memory Usage vs Embedding Size}
  \label{fig:MemvsEmb}
\end{figure*}

\begin{figure*}
  \centering
  \includegraphics[width=\textwidth]{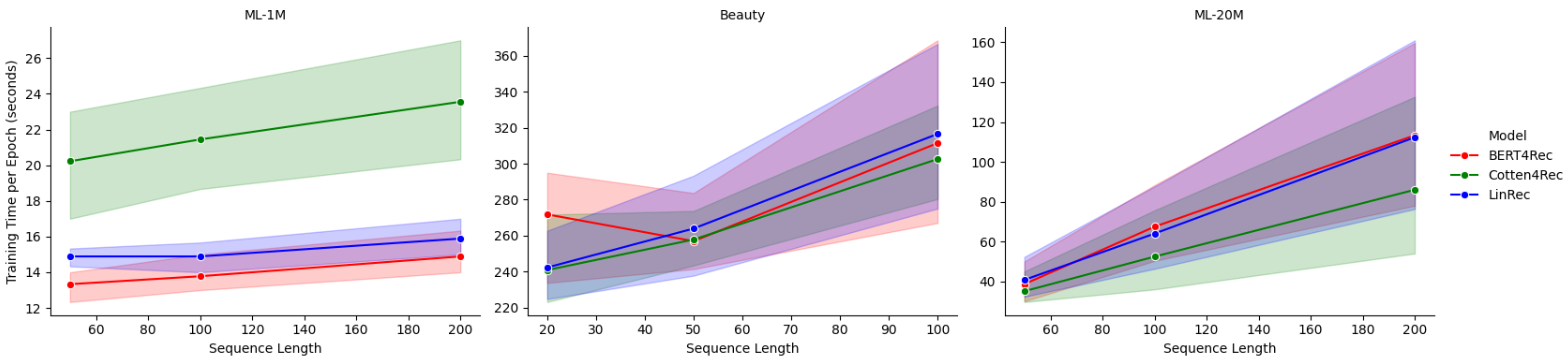}
  \caption{Training Time per Epoch vs Sequence Length}
  \label{fig:TTvsSeqL}
\end{figure*}

\begin{figure*}
  \centering
  \includegraphics[width=\textwidth]{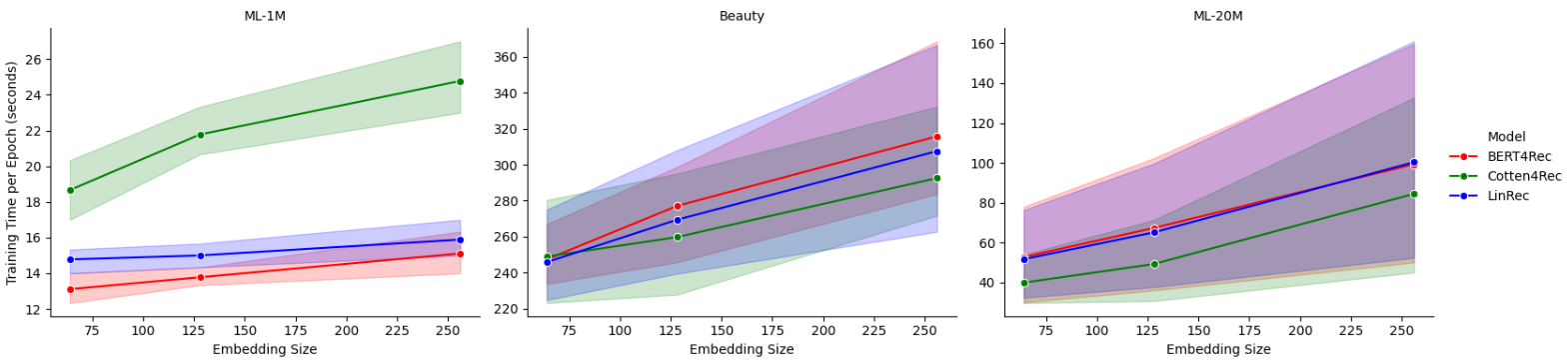}
  \caption{Training Time per Epoch vs Embedding Size}
  \label{fig:TTvsEmb}
\end{figure*}

Across all three benchmark datasets, Cotten4Rec cuts peak GPU memory by $\approx23\%$ compared to both BERT4Rec and LinRec. On Beauty and ML-20M, it also accelerates training by $\approx4\%$ and $\approx20\%$, respectively, while keeping NDCG@10 and HIT@10 within 2\% of the baselines. In contrast, on ML-1M, Cotten4Rec trains $\approx49\%$ slower and sees an $\approx11\%$ drop in NDCG@10 and a $\approx8\%$ drop in HIT@10 versus BERT4Rec and LinRec. Figures \ref{fig:MemvsSeqL}-\ref{fig:TTvsEmb} display the memory and runtime trends of Tables \ref{tab:cotten4rec-comparison-sequencelength} and \ref{tab:cotten4rec-comparison-embeddingsize}, which present metrics on the performance and efficiency of each model for each dataset across varying sequence lengths and embedding sizes.

1

\begin{table*}
  \caption{Comparison of Cotten4Rec Performance Across various Sequence Lengths with Bert4Rec and LinRec Models}
  \label{tab:cotten4rec-comparison-sequencelength}
  \begin{tabular}{l c c c c c c c c c c}
    \toprule
    Dataset & SeqL & Model & Mem(MB) & Training(s) & NDCG@10 & HIT@10 & MB(\%) & Time(\%) & NDCG(\%) & HIT(\%) \\
    \midrule
    Beauty & 20 & BERT4Rec & 5246 & 1359 & 0.249 & 0.395 & -21.96\% & -11.42\% & -1.76\% & -0.77\% \\
     &  & LinRec & 5269 & 1210 & 0.248 & 0.394 & -22.29\% & -0.52\% & -1.48\% & -0.49\% \\
     &  & Cotten4Rec & 4094 & 1204 & 0.244 & 0.392 &  &  &  &  \\
     & 50 & BERT4Rec & 12473 & 1285 & 0.250 & 0.396 & -23.61\% & 0.30\% & -2.16\% & -0.98\% \\
     &  & LinRec & 12499 & 1319 & 0.248 & 0.393 & -23.77\% & -2.35\% & -1.33\% & -0.17\% \\
     &  & Cotten4Rec & 9528 & 1288 & 0.245 & 0.392 &  &  &  &  \\
     & 100 & BERT4Rec & 24521 & 1556 & 0.248 & 0.392 & -24.24\% & -2.78\% & -1.03\% & 0.01\% \\
     &  & LinRec & 24534 & 1583 & 0.248 & 0.393 & -24.28\% & -4.38\% & -1.07\% & -0.24\% \\
     &  & Cotten4Rec & 18578 & 1513 & 0.246 & 0.392 &  &  &  &  \\
    \midrule
    ML-1M & 50 & BERT4Rec & 667 & 266 & 0.381 & 0.609 & -19.02\% & 51.91\% & -16.75\% & -11.23\% \\
     &  & LinRec & 692 & 297 & 0.357 & 0.588 & -21.93\% & 36.09\% & -11.11\% & -7.97\% \\
     &  & Cotten4Rec & 540 & 404 & 0.318 & 0.541 &  &  &  &  \\
     & 100 & BERT4Rec & 1318 & 273 & 0.371 & 0.602 & -21.06\% & 56.39\% & -15.43\% & -10.04\% \\
     &  & LinRec & 1332 & 296 & 0.346 & 0.580 & -21.88\% & 44.25\% & -9.13\% & -6.58\% \\
     &  & Cotten4Rec & 1040 & 427 & 0.314 & 0.542 &  &  &  &  \\
     & 200 & BERT4Rec & 2689 & 298 & 0.358 & 0.593 & -24.13\% & 58.51\% & -13.38\% & -9.26\% \\
     &  & LinRec & 2612 & 316 & 0.322 & 0.561 & -21.91\% & 49.18\% & -3.84\% & -4.07\% \\
     &  & Cotten4Rec & 2040 & 472 & 0.310 & 0.538 &  &  &  &  \\
    \midrule
    ML-20M & 50 & BERT4Rec & 2998 & 777 & 0.750 & 0.932 & -22.77\% & -9.41\% & -1.81\% & -0.55\% \\
     &  & LinRec & 3023 & 818 & 0.749 & 0.932 & -23.41\% & -13.93\% & -1.62\% & -0.55\% \\
     &  & Cotten4Rec & 2315 & 704 & 0.737 & 0.927 &  &  &  &  \\
     & 100 & BERT4Rec & 5895 & 1350 & 0.755 & 0.936 & -23.64\% & -22.21\% & -2.12\% & -0.57\% \\
     &  & LinRec & 5909 & 1279 & 0.752 & 0.936 & -23.82\% & -17.90\% & -1.75\% & -0.60\% \\
     &  & Cotten4Rec & 4501 & 1050 & 0.739 & 0.931 &  &  &  &  \\
     & 200 & BERT4Rec & 11769 & 2267 & 0.756 & 0.938 & -24.58\% & -24.06\% & -2.44\% & -0.54\% \\
     &  & LinRec & 11694 & 2247 & 0.750 & 0.936 & -24.09\% & -23.39\% & -1.61\% & -0.39\% \\
     &  & Cotten4Rec & 8877 & 1722 & 0.738 & 0.933 &  &  &  &  \\
    \bottomrule
    \end{tabular}
\end{table*}

\begin{table*}[t]
  \caption{Comparison of Cotten4Rec Performance Across various Embedding Sizes with Bert4Rec and LinRec Models}
  \label{tab:cotten4rec-comparison-embeddingsize}
  \begin{tabular}{l c c c c c c c c c c}
    \toprule
    Dataset & Embedding & Model & Mem(MB) & Training(s) & NDCG@10 & HIT@10 & MB(\%) & Time(\%) & NDCG(\%) & HIT(\%) \\
    \midrule
    Beauty & 64  & BERT4Rec     & 13675 & 1237 & 0.25 & 0.40 & -24.51\% & 0.58\%  & -0.05\% & 0.08\% \\
           &     & LinRec       & 13667 & 1228 & 0.25 & 0.40 & -24.46\% & 1.29\%  & -0.33\% & -0.38\% \\
           &     & Cotten4Rec   & 10324 & 1244 & 0.25 & 0.40 &          &         &         &        \\
     & 128 & BERT4Rec     & 13978 & 1384 & 0.25 & 0.40 & -23.95\% & -6.17\% & -2.46\% & -1.44\% \\
           &     & LinRec       & 13989 & 1346 & 0.25 & 0.40 & -24.01\% & -3.48\% & -1.92\% & -0.76\% \\
           &     & Cotten4Rec   & 10630 & 1299 & 0.25 & 0.39 &          &         &         &        \\
     & 256 & BERT4Rec     & 14588 & 1579 & 0.24 & 0.38 & -22.91\% & -7.38\% & -2.45\% & -0.39\% \\
           &     & LinRec       & 14646 & 1538 & 0.24 & 0.38 & -23.21\% & -4.92\% & -1.66\% & 0.26\%  \\
           &     & Cotten4Rec   & 11247 & 1463 & 0.24 & 0.38 &          &         &         &        \\
    \midrule
    ML-1M  & 64  & BERT4Rec     & 1208  & 261  & 0.27 & 0.48 & -26.39\% & 42.99\% & -2.47\% & -2.12\% \\
           &     & LinRec       & 1151  & 294  & 0.26 & 0.48 & -22.70\% & 27.16\% & -1.17\% & -1.14\% \\
           &     & Cotten4Rec   & 889   & 374  & 0.26 & 0.47 &          &         &         &        \\
      & 128 & BERT4Rec     & 1467  & 275  & 0.37 & 0.62 & -23.36\% & 58.02\% & -18.63\% & -14.48\% \\
           &     & LinRec       & 1440  & 299  & 0.33 & 0.58 & -21.91\% & 45.67\% & -8.65\%  & -8.38\%  \\
           &     & Cotten4Rec   & 1124  & 435  & 0.30 & 0.53 &          &         &         &        \\
      & 256 & BERT4Rec     & 1998  & 300  & 0.48 & 0.71 & -19.61\% & 64.70\% & -19.72\% & -11.93\% \\
           &     & LinRec       & 2046  & 317  & 0.43 & 0.68 & -21.46\% & 56.05\% & -12.02\% & -8.01\%  \\
           &     & Cotten4Rec   & 1607  & 494  & 0.38 & 0.62 &          &         &         &        \\
    \midrule
    ML-20M & 64  & BERT4Rec     & 6491  & 1059 & 0.72 & 0.92 & -25.08\% & -24.41\% & -3.08\% & -0.86\% \\
           &     & LinRec       & 6435  & 1032 & 0.72 & 0.92 & -24.42\% & -22.47\% & -2.35\% & -0.86\% \\
           &     & Cotten4Rec   & 4863  & 800  & 0.70 & 0.92 &          &         &         &        \\
     & 128 & BERT4Rec     & 6787  & 1346 & 0.76 & 0.94 & -24.26\% & -26.85\% & -2.19\% & -0.60\% \\
           &     & LinRec       & 6761  & 1304 & 0.75 & 0.94 & -23.97\% & -24.53\% & -1.56\% & -0.51\% \\
           &     & Cotten4Rec   & 5140  & 984  & 0.74 & 0.93 &          &         &         &        \\
     & 256 & BERT4Rec     & 7383  & 1989 & 0.78 & 0.94 & -22.94\% & -15.01\% & -1.17\% & -0.21\% \\
           &     & LinRec       & 7429  & 2007 & 0.78 & 0.94 & -23.42\% & -15.76\% & -1.11\% & -0.17\% \\
           &     & Cotten4Rec   & 5689  & 1691 & 0.77 & 0.94 &          &         &         &        \\
    \bottomrule
  \end{tabular}
\end{table*}

\section{Discussion}

\subsection*{A. Memory Usage}

Cotten4Rec achieves a $\approx23\%$ reduction in peak GPU memory by eliminating the need to store the full $s\cdot s$ attention score matrix (and its gradient) in each Transformer layer.  In BERT4Rec, the scaled dot‐product attention computes
\[
\mathrm{scores} = QK^\top \quad\bigl(s\cdot  d \;\times\; d\cdot  s \to s\cdot  s\bigr)
\]
and retains this $s^2$ buffer for the backward pass, costing $O(s^2 d)$ memory per head.  LinRec also materializes per‐head feature maps of shape $s\cdot  d$ and additional intermediate buffers for its ELU‐based reformulation, leading to the same $O(s^2 d)$ memory footprint in practice.  

By contrast, Cotten4Rec rewrites the attention as
\[
K^\top V \;\bigl(d\cdot  s\;\times\; s\cdot  d \to d\cdot  d\bigr)
\;\;\xrightarrow{\quad}\;\;
Q \,(K^\top V)\;\bigl(s\cdot  d\;\times\; d\cdot  d \to s\cdot  d\bigr)
\]
so the largest temporary activation is only $d\cdot  d$, i.e.\ $O(d^2)$ per head.  Both LinRec and Cotten4Rec share the same asymptotic memory and compute complexity $O(s\cdot d^2)$, but LinRec’s reference implementation issues separate PyTorch kernels for ELU, normalization, and batched GEMMs, each allocating extra scratch space and incurring launch overhead.  Our custom CUDA kernel fuses the $\ell_2$‐normalization, matrix multiplications, and reduction steps into a single, memory‐coalesced operation, eliminating those temporaries.  This fusion not only realizes the theoretical $O(s\cdot d^2)$ savings but also delivers a consistent constant‐factor speedup over LinRec’s more fragmented pipeline.

\subsection*{B. Training Time}

Training runtime depends on the relative cost of attention versus embedding operations, which varies with sequence length $s$ and vocabulary size $|\mathcal I|$.  On Beauty ($s\approx9$, $|\mathcal I|=120$K) and ML-20M ($s\approx68$, $|\mathcal I|=16.6$K), attention comprises only a small portion of the per-epoch workload.  In these regimes, Cotten4Rec’s $O(s\cdot d^2)$ cosine‐attention kernel—implemented as a single fused CUDA operation—outperforms both BERT4Rec’s Softmax attention and LinRec’s multi-kernel ELU/GEMM pipeline, yielding a modest $\approx4\%$ speedup on Beauty and a larger $\approx20\%$ gain on ML-20M.

In contrast, ML-1M’s long histories ($s\approx166$, $|\mathcal I|=3.7$K) shift the bottleneck to the attention computation itself.  Here, BERT4Rec’s single, highly-optimized Softmax kernel (and LinRec’s fused ELU/GEMM) incur lower constant-factor overhead than Cotten4Rec’s normalization and split‐kernel approach, making Cotten4Rec $\approx49\%$ slower than BERT4Rec and LinRec.

Embedding cost scales linearly in $|\mathcal I|\times d$ and does not affect attention complexity.  Thus, Beauty’s very large vocabulary amplifies embedding overhead, limiting the relative benefit from faster attention, whereas ML-1M’s small vocabulary shifts the performance balance entirely onto the attention kernel.  These empirical findings align with theoretical expectations for $O(s\cdot d^2)$ versus $O(s^2\cdot  d)$ attention under different $(s,|\mathcal I|)$ regimes.  

\subsection*{C. Recommendation Quality}

Cotten4Rec replaces the precise Softmax weighting with cosine‐similarity scores, which effectively smooths the attention distribution.  As a result, it cannot distinguish subtle differences between many candidate items as sharply, leading to consistently lower NDCG@10 and HIT@10 compared to both BERT4Rec and LinRec.  

When user histories are short or moderate (Beauty, ML-20M), the top‐ranked items overwhelmingly dominate, so this smoothing costs under 2\% in NDCG@10 and under 1\% in HIT@10.  For very long histories (ML-1M, $s\approx166$), however, Softmax’s ability to leverage small score differences across many positions becomes critical; the cosine approximation’s averaging effect compounds across $s$, producing an $\approx11\%$ drop in NDCG@10 and a $\approx8\%$ drop in HIT@10.  

Thus, while cosine attention offers major efficiency gains, it trades off ranking precision on datasets with long, information‐rich sequences.

\section{Conclusion, Limitations, and Future Work}

\subsection*{A. Conclusion}
In this research, we introduced Cotten4Rec, a novel sequential‐recommendation model that replaces Softmax‐scaled dot‐product attention with an $L_{2}$-normalized cosine similarity attention and optimized CUDA kernel. Against the state-of-the-art BERT4Rec and the linear-attention baseline LinRec, Cotten4Rec consistently reduces peak GPU memory by $\approx23\%$ on three real-world datasets: Amazon Beauty, MovieLens-1M, and MovieLens-20M.  It further accelerates training by $\approx4\%$ on short histories and $\approx20\%$ on moderate-length histories, while maintaining NDCG@10 and HIT@10 within $2\%$ of the baselines.  These results demonstrate that cosine attention, when fused into a high-throughput CUDA kernel, provides a compelling efficiency–accuracy trade-off for large-vocabulary, short-to-medium sequence SR tasks.

\subsection*{B. Limitations}
Despite its efficiency gains, Cotten4Rec exhibits two key limitations.  First, on very long sequences (e.g.\ ML-1M, $s\approx166$), the simplified cosine kernel incurs higher constant-factor overhead than a mature Softmax or ELU/GEMM pipeline, resulting in $\approx49\%$ slower training.  Second, the smoothing effect of cosine similarity degrades recommendation quality more substantially as sequence length increases, with up to $15\%$ loss in NDCG@10 and a $10\%$ drop in HIT@10.  Moreover, the dependence on a custom CUDA kernel may complicate portability to other hardware or software environments.

\subsection*{C. Future Work}

Future research can expand on this foundation by conducting large‐scale ablation studies over embedding dimension, sequence length, mask probability, and dropout across additional domains (e.g.\ e‐commerce, streaming) to assess generalizability. Integrating cosine attention into decoder‐only Transformer architectures (e.g.\ GPT‐style) could reveal its impact in an autoregressive setting and clarify encoder versus decoder-based trade-offs. Investigating adaptive attention fusion—dynamically switching between cosine and Softmax attention at runtime based on sequence length, embedding size, or hardware characteristics—may optimize performance and accuracy. Finally, exploring low-rank pruning or quantization techniques for the embedding table could further reduce memory and compute costs when combined with cosine attention.

\section*{Acknowledgements}
The authors thank Advait Pavuluri (undergraduate student at RPI) for their assistance with the experimental runs.

%
%

\end{document}